\documentclass[11pt,reqno]{amsart}
\usepackage[utf8]{inputenc}
\usepackage[top=1in, bottom=1in, left=1in, right=1in]{geometry}

\usepackage{amsmath}
\usepackage{amsfonts}
\usepackage{amssymb}
\usepackage{amsthm}
\numberwithin{equation}{section}

\theoremstyle{plain}
\newtheorem{thm}{Theorem}[section]
\newtheorem{theorem}[thm]{Theorem}

\newtheorem{prop}[thm]{Proposition}
\newtheorem{remark}{Remark}

\usepackage{tikz}
\usetikzlibrary{
    shapes,
    decorations,
    arrows,
    calc,
    arrows.meta,
    fit,
    positioning,
    matrix 
    }
\tikzset{
    > = stealth,
    bidirected/.append style={
        arrows = <->,
        dashed,
        },
    }

\usepackage{hyperref}
\hypersetup{colorlinks}

\def \bM{\mathbf{M}}
\def \bC{\mathbf{C}}
\def \bS{\mathbf{S}}
\def \bF{\mathbf{F}}
\def \bbeta{\boldsymbol{\beta}}
\def \balpha{\boldsymbol{\alpha}}
\def \bs{\mathbf{s}}
\def \bf{\mathbf{f}}

\def \tM{\widetilde{M}}
\def \dM{\dot{M}}

\def \lp{\mathrm{L}}

\def \hm{\mathbf{m}}
\def \hp{\mathbf{p}}

\def \E{\mathbb{E}}
\def \Pb{\mathbb{P}}
\def \Var{\mathrm{Var}}
\def \Cov{\mathrm{Cov}}

\def \spdot{\; \cdot}

\newcommand{\indep}{\perp \!\!\! \perp}
\newcommand{\dep}{\not\!\perp\!\!\!\perp}

\usepackage{amsaddr}
\author{Hanbin Lee$^{\dagger,\ddagger}$ and Moohyuk Lee$^\dagger$}
\address{
    $^{\dagger}$Department of Medicine, Seoul National University College of Medicine\\Seoul, Republic of Korea\\
    $^{\ddagger}$Department of Mathematical Sciences, Seoul National University\\Seoul, Republic of Korea
    }
\email{hanbin973@snu.ac.kr}

\title{Disentangling Linkage and Population Structure in Association Mapping}

\begin{document}

\maketitle

\begin{abstract}
   Genome-wide association study (GWAS) tests single nucleotide polymorphism (SNP) markers across the genome to localize the underlying causal variant of a trait.
   Because causal variants are seldom observed directly, a surrogate model based on genotyped markers are widely considered.
   Although many methods estimating the parameters of the surrogate model have been proposed, the connection between the surrogate model and the true causal model is yet investigated.
   In this work, we establish the connection between the surrogate model and the true causal model.
   The connection shows that population structure is accounted in GWAS by modelling the variant of interest and not the trait.
   Such observation explains how environmental confounding can be partially corrected using genetic covariates and why the previously claimed connection between PC correction and linear mixed models is incorrect.
\end{abstract}

\section{Introduction}
\textit{Genome-wide association study} (GWAS) identifies regions in the genome responsible for the variation in a trait through \textit{single nucleotide polymorphisms} (SNPs) distributed across the genome.
Each SNP is tested for its association with the trait (generally through regression models) and its regression coefficient (henceforth, marker effect size) is recorded.  
Although markers themselves may not be causal, a significant effect size hints a true causal variant nearby the marker being tested.
This is due to linkage, the correlation between two physically proximal variants.
Earlier studies have found that SNPs are dense and widespread across the genome \cite{2005, 2007}.
Therefore, testing SNPs across the genome enables the identification of genetic signals genome-wide even if some causal variants are missing which gives the name of GWAS.

For continuous traits, the \textit{quantitative trait model} is widely adopted where the trait is an additive function respect to the observed SNPs.
From now on, we call this model as the \textit{marker-additive model} (MAM).
The estimation of MAM parameters have been extensively studied and many methods have been proposed \cite{Price_2006, Kang_2010, Mbatchou_2021}.
MAM, however, is a surrogate model where the parameters (the marker effect sizes) have no direct causal interpretation since SNP markers are seldom causal themselves.
Furthermore, both empirical and theoretical work have shown that stratification of allele frequencies due to population structure can produce spurious signals without linkage \cite{Rosenberg_2006}.

In this work, we consider an additive model respect to the causal variants, the \textit{causal-additive model} (CAM), and establish the connection with the MAM.
CAM parameters have direct causal interpretation because it directly involves often unobserved causal variants of the trait.
We deliver a formula showing that marker effect size can be decomposed into three terms that reflect linkage and population structure separately.
Best to our knowledge, our work is the first attempt to address the identification problem of covariate adjusted regression in GWAS.

\section{Setup}
\subsection{Notations and Models}
The data discussed in this work is $(Y_i, \bC_i, \bM_i, \bS_i)_{i=1}^n$ where $Y_i \in \mathbb{R}$ is the trait, $\bC_i \in \mathbb{R}^p$ is the causal variants, $\bM_i \in \mathbb{R}^q$ is the markers and $\bS_i$ is the population membership. 
In this study, we focus on diploid populations where elements of $\bC_i$ and $\bM_i$ take values $0,1,2$.
$h=\hm,\hp$ denotes the haplotype index ($\hm$ for maternal and $\hp$ for paternal haplotype) so that $C_{ijh}=0,1$ and $M_{ikh}=0,1$, accordingly.
The \textit{marker-additive model} (MAM) is

\begin{align}
    Y_i = \bM_i^T\bbeta + \beta_0 + \delta_i \text{\quad and\quad} \E[\delta_i \mid \bM_i]=0 
    \label{eq:marker_whole}
\end{align}
where $\bbeta \in \mathbb{R}^q $ is the marker effect size and $\delta_i$ is the noise variable.
In most cases with very few exceptions, markers are tested marginally rather than as a whole.
Without loss of generality, we write $M_{i1}$ as the marker being tested which leads to the following regression.

\begin{align}
    Y_i = M_{i1} \beta_1 + \bM_{i(-1)}^T \bbeta_{(-1)} + \beta_0 + \delta_i
    \label{eq:marker_marginal}
\end{align}
where $(-1)$ in the subscript means that the first element is dropped from the original variable.
Although model (\ref{eq:marker_whole}) contains all genotyped markers, in marginal testing, markers that are physically close to the variant being tested are removed from the regression.
One such way is the \textit{leave one chromosome out} (LOCO) approach where markers appearing in $\bM_{i(-1)}$ are all sampled outside the chromosome which contains the variant being tested \cite{Yang_2014}.

The \textit{causal-additive model} (CAM), analogous to MAM, is

\begin{align}
    Y_i = \bC_i^T \balpha + \alpha_0 + \epsilon_i \text{\quad and\quad} \E[\epsilon_i \mid \bC_i]=0
    \label{eq:causal_whole}
\end{align}
where $\balpha \in \mathbb{R}^p$ is the causal effect size and $\epsilon_i$ is the noise variable.
Since changes in $\bC_i$ has direct influence on $Y_i$, $\balpha$ has a straightforward causal interpretation.

Several population structure models are discussed in this work.
When the sample is obtained from $L$ discrete populations indexed by $l$, $\bS_i \in \mathbb{R}^{L}$ where

\begin{gather*}
    \begin{gathered}
    \bS_i =
        (0, \ldots, 0, \ldots, 0) \text{ if the $i$-th individual is from population } l=1\\
        \text{or} \\
        \bS_i = (0, \ldots, \underset{\underset{l-\text{th}}{\uparrow}}{1}, \ldots, 0) \text{ if the $i$-th individual is from population } l \neq 1
        \end{gathered}
\end{gather*}
The \textit{Pritchard-Stephens-Donnelly} (PSD) model is a more general model that describes population structure incorporating \textit{admixture}, where the genetic materials of an individual originates from multiple ancestries (indexed by $l$ with total $L$ of them) rather than a single one \cite{Pritchard_2000}.
In this case, $\bS_i$ is often called the \textit{admixture proportion} where

\begin{gather*}
    \bS_i = (a_{i1}, \ldots, a_{il}, \ldots, a_{iL})
\end{gather*}
with $\sum_{l=1}^L a_{il} =1$.
The $l$-th element is the proportion of genome of an individual that originates from the $l$-th ancestral population.
If one restricts the elements of $\bS_i$ to take values $0$ or $1$, the PSD model reduces to the discrete populations model.
The PSD model further assumes that the \textit{Hardy-Weinberg Equilibrium} (HWE) holds conditional on $\bS_i$.
That being said, random mating is assumed and as a consequence, an individual receives its genome by randomly selecting a haplotype from the parent generation.

We define haplotype frequencies and \textit{linkage disequilibrium} (LD) parameters.
$f_j = \Pb(C_{ijh}=1)$, $g_k = \Pb(M_{ikh}=1)$ and $h_{jk} = \Pb(C_{ijh}=1, M_{ikh}=1)$ are allele frequencies and joint allele frequency of causal variant $j$, marker variant $k$ and both variant $j$ and variant $k$.
LD covariance $D_{jk} = h_{jk} - f_jh_k$ is the covariance of variant $j$ and $k$.
When considering population membership, population specific analogues should be defined.
$f_j^{\bs} = \Pb(C_{ijh}=1 \mid \bS_i=\bs)$, $g_k^{\bs} = \Pb(M_{ikh}=1 \mid \bS_i=\bs)$ and $h_{jk}^{\bs} = \Pb(C_{ijh}=1, M_{ikh}=1 \mid \bS_i=\bs)$ are the population specific allele frequencies and joint allele frequency.
Population specific LD covariance is defined accordingly $D_{jk}^{\bs} = h_{jk}^{\bs} - f_j^{\bs} h_k^{\bs}$.

Finally, linear projection of $X$ respect to $Y$, $\lp[X \mid Y]$, will be frequently used throughout \cite{Wooldridge_2010}.
Note that the variables appearing on the right side the vertical sign ($\vert$) only contain the random components, and the intercept $1$ is always included but ignored in the notation for convenience.

\subsection{The Causal Structure}\label{subsec:cstruct}

The conditional independence implied by the causal structure plays a pivotal role in this work.
The causal structure is summarized in the following causal diagram.

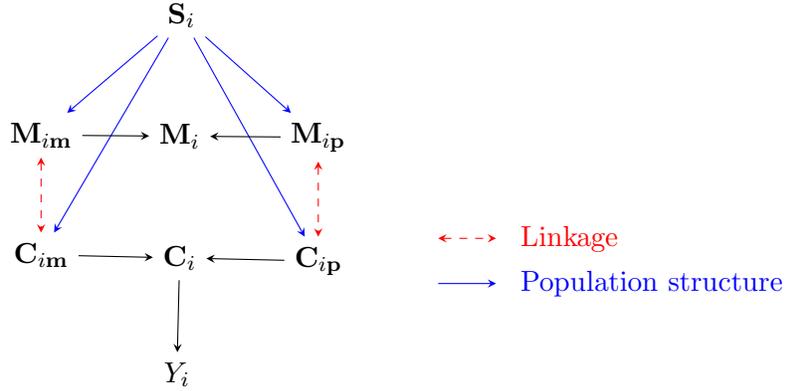
\begin{figure}[ht]
    \centering
    \begin{tikzpicture}
        \node (mm) {$\bM_{i\hm}$};
        \node (s) [above right = of mm] {$\bS_i$};
        \node (mp) [below right = of s] {$\bM_{i\hp}$};
        \node (cm) [below = of mm] {$\bC_{i\hm}$};
        \node (cp) [below = of mp] {$\bC_{i\hp}$};
        \node (y) [below right = of cm] {$Y_i$};
        
        \node (m) at ($(mp)!0.5!(mm)$) {$\bM_i$};
        \node (c) at ($(cp)!0.5!(cm)$) {$\bC_i$};
        
        \path[->,blue] (s) edge (mm);
        \path[->,blue] (s) edge (mp);
        \path[->,blue] (s) edge (cm);
        \path[->,blue] (s) edge (cp);
        
        \path[->] (mm) edge (m);
        \path[->] (mp) edge (m);
        \path[->] (cm) edge (c);
        \path[->] (cp) edge (c);
        \path[->] (c) edge (y);
        
        \path[bidirected, red] (mm) edge (cm);
        \path[bidirected, red] (mp) edge (cp);
        
        \matrix [column sep=0.5em, right = of cp] {
            \draw[bidirected,red] (0em,0em) -- (2em,0em); & \node[red] {Linkage}; \\
            \draw[->,blue] (0em,0em) -- (2em,0em); & \node[blue] {Population structure}; \\
        };
    \end{tikzpicture}
    \label{dag:main}
\end{figure}
The key condition we later use implied by the diagram is $\bM_{i\hm}, \bC_{i\hm} \indep \bM_{i\hp}, \bC_{i\hp} \mid \bS_i$.
In other words, the genotype at different haplotypes are conditionally independent which is also implied by the conditional HWE assumption.

The diagram shows the two sources of association between the marker ($\bM_i$) and the trait ($Y_i$):

\begin{enumerate}
    \item 
    \tikz[baseline]{
        \node[anchor=base] (m) {$\bM_{ih}$}; 
        \node[anchor=base] (c) [right = of m] {$\bC_{ih}$}; 
        \node[anchor=base] (y) [right = of c] {$Y_i$}; 
        \path[bidirected, red] (c) edge (m); 
        \path[->] (c) edge (y);
        }: 
    Association due to linkage.
    \item 
    \tikz[baseline]{
        \node[anchor=base] (m) {$\bM_{ih}$}; 
        \node[anchor=base] (s) [right = of m]{$\bS_i$};
        \node[anchor=base] (c) [right = of s] {$\bC_{ih}$}; 
        \node[anchor=base] (y) [right = of c] {$Y_i$}; 
        
        \path[->, blue] (s) edge (c); 
        \path[->, blue] (s) edge (m);
        \path[->] (c) edge (y);
        }: 
    Association due to population structure.
\end{enumerate}
In GWAS, we only want the first association to be present.
This can be achieved by conditioning on $\bS_i$, but it is generally not observed in population samples.
We later show that this problem is partially, but not fully, accounted by including many markers to the regression as in (\ref{eq:marker_marginal}).

\subsection{The Goal}
Our goal is to characterize the estimand of the regression (\ref{eq:marker_marginal}) under the CAM (\ref{eq:causal_whole}).
Following the practice of excluding variants that are near to the marker being tested, we assume that $M_{i1} \indep \bM_{i(-1)} \mid \bS_i$.
Due to the causal structure, this is equivalent to saying that $M_{i1}$ is unlinked to the rest of the variants in the regression.
Hence, conditional on $\bS_i$, the following regression and regression (\ref{eq:marker_marginal}) has the same estimand.

\begin{align}
    Y_i = M_{i1} \beta_{1,\mathrm{nocov}} + \beta_{0,\mathrm{nocov}} +  \delta_{i, \mathrm{nocov}}
    \label{eq:marker_marginal_nocov}
\end{align}
i.e. $\beta_1 = \beta_{1,\mathrm{nocov}}$ when the regression is restricted to a fixed $\bS_i$.
Because the regression is conditioned on $\bS_i$, $\beta_1$ only reflects route (1) in section (\ref{subsec:cstruct}).
To distinguish it from the regression coefficient obtained from the whole sample without restricting $\bS_i$, we write this estimand as $\beta_1(\bS_i)$.
It can be shown that 

\begin{align}
    \begin{aligned}
        \beta_1(\bS_i = \bs) 
        &= \sum_j \frac{D_{j1}^{\bs}}{g_1^{\bs} (1-g_1^{\bs})} \alpha_j \\
        &= \sum_j \beta_{1j}(\bS_i = \bs) 
    \end{aligned}
\end{align}
by the formula of univariate linear regression.
Given that $\E[C_{ij} \mid \bS_i]$ and $\E[M_{ik} \mid \bS_i]$ are linear functions respect to $\bS_i$ (e.g. admixture), the regression applied to the whole population 

\begin{align*}
    Y_i = M_{1i} \beta_{1,\bs} + \bS_i \boldsymbol{\gamma}_{\bs} + \beta_{0,\bs} + \delta_{i,\bs}
\end{align*}
has the estimand

\begin{align}
    \begin{aligned}
        \beta_{1,\bs} &= \sum_j \frac{\E_{\bS}[\beta_{1j}(\bS_i) \Var(M_{i1} \mid \bS_i)]}{\E_{\bS}[\Var(M_{i1} \mid \bS_i)]} \\
        &= \frac{\E_{\bS}[\beta_{1}(\bS_i) \Var(M_{i1} \mid \bS_i)]}{\E_{\bS}[\Var(M_{i1} \mid \bS_i)]}
    \end{aligned}
    \label{eq:marker_marginal_s}
\end{align}
similar to the case of OLS under \textit{heterogeneous treatment effect} (HTE) \cite{Imbens_2009}.
Equation (\ref{eq:marker_marginal_s}) is a weighted average over $\beta_1(\bS_i)$ with weights that sum up to $1$.

In the whole sample, two regressions (\ref{eq:marker_marginal}) and (\ref{eq:marker_marginal_nocov}) have different estimands because $M_{i1} \dep \bM_{i(-1)}$ in general.
We aim to understand how $\beta_1$, 
the marker effect size obtained from regression (\ref{eq:marker_marginal}) applied to the whole population, 
is related to $\beta_1(\bS_i)$, the desired estimand obtained from ideal populations each in HWE.

\section{Main Results}

In GWAS, we are only interested in the contribution of linkage with a physically proximal causal variant.
It should be desirable if path (1) in section (\ref{subsec:cstruct}) can be separated from path (2).
However, we show that this is only partially achievable under the population-based design.
Nevertheless, it is achievable under the within-sibship design \cite{Howe_2022}.

\subsection{Population-based GWAS}
We begin by stating the result for a simple regression (\ref{eq:marker_marginal_nocov}) without any covariates.
\begin{prop}
The estimand of regression (\ref{eq:marker_marginal_nocov}) on the whole population is

\begin{align*}
    \begin{aligned}
        \beta_{1, \mathrm{nocov}} 
        &= \sum_j \underbrace{\frac{\E_{\bS}[\beta_{1j}(\bS_i)\Var(M_{i1} \mid \bS_i)]}{\Var[M_{i1}]}}_{\text{linkage}}
        + \sum_j \underbrace{\frac{\Cov_{\bS}[\E(C_{ij} \mid \bS_i), \E(M_{i1} \mid \bS_i)] }{\Var[M_{i1}]} \cdot \alpha_j}_{\text{population structure}} \\
        &= \frac{\E_{\bS}[\beta_{1}(\bS_i)\Var(M_{i1} \mid \bS_i)]}{\Var[M_{i1}]}
        + \sum_j \frac{\Cov_{\bS}[\E(C_{ij} \mid \bS_i), \E(M_{i1} \mid \bS_i)] }{\Var[M_{i1}]} \cdot \alpha_j
    \end{aligned}
\end{align*}
\label{prop:marker_marginal_nocov}
\end{prop}

\begin{remark}[The Wahlund effect]
The first term can be viewed as a average of $\beta_1(\bS_i)$ weighted by $\frac{\Var(M_{i1} \mid \bS_i)\Pb(\bS_i)}{\Var[M_{i1}]}$.
An important observation is that the weights sum up to a value smaller than $1$ as long as there is allele frequency stratification of $M_{i1}$ across $\bS_i$.
To see this,

\begin{align*}
    \Var[M_{i1}] 
    = \E_{\bS}[\Var(M_{i1} \mid \bS_i)] + \Var_{\bS}[\E_{\bS}(M_{i1} \mid \bS_i)]
\end{align*}
which is due to the law of total variance.
This was first reported in 1928 by Wahlund \cite{WAHLUND_2010}.
The sum over the weights can also be expressed in terms of Wright's $F$-statistics \cite{Wright_1943} which is a popular measure of population structure.

\begin{align*}
    \frac{\E_{\bS}[\Var(M_{i1} \mid \bS_i)]}{\Var[M_{i1}]} = 1-F_{\mathrm{ST}}
\end{align*}

A straightforward consequence of the Wahlund effect is that the desired effect $\beta_1(\bS_i)$ is attenuated and the attenuation is proportional to the strength of population structure, namely, $F_{\mathrm{ST}}$.

\end{remark}

The second term due to population structure in \textbf{Proposition \ref{prop:marker_marginal_nocov}} has a more familiar interpretation.
The numerator is non-zero if and only if the causal variant and the marker variant allele frequencies align simultaneously across $\bS_i$.
In sum, \textbf{Proposition \ref{prop:marker_marginal_nocov}} shows that population structure affects $\beta_{1, \mathrm{nocov}}$ in two folds.
It attenuates the true signal and puts undesirable signals into the estimand.

Now we propose our main theorem which delivers the estimand of regression (\ref{eq:marker_marginal}).
\begin{theorem}
    Let $\tM_{i1} = M_{i1} - \lp[M_{i1} \mid \bM_{i(-1)}]$.
    The estimand of regression (\ref{eq:marker_marginal}) is
    
    \begin{align*}
        \begin{aligned}
            \beta_1 
            &= \sum_j \underbrace{\frac{\E_{\bS}[\beta_{1j}(\bS_i)\Var(M_{i1} \mid \bS_i)]}{\Var[\tM_{i1}]}}_{\text{linkage}}
            \;\left(=\frac{\E_{\bS}[\beta_{1}(\bS_i)\Var(M_{i1} \mid \bS_i)]}{\Var[\tM_{i1}]} \right) 
            \\
            &+ 
            \sum_j \underbrace{\frac{
                \E[C_{ij} ( \E[M_{i1} \mid \bS_i] - \E_{\bS}[\E(M_{i1} \mid \bS_i) \mid \bM_{i(-1)}])]
                }{
                    \Var[\tM_{i1}] 
                } \cdot \alpha_j
            }_{\text{prediction error}} \\
            &+ \sum_j \underbrace{\frac{
                \E[C_{ij} ( \E[M_{i1} \mid \bM_{i(-1)}] - \lp[M_{i1} \mid \bM_{i(-1)}])] 
                }{
                    \Var[\tM_{i1}] 
                } \cdot \alpha_j
            }_{\text{functional misspecification}} 
        \end{aligned}
    \end{align*}
    \label{thm:marker_marginal}
\end{theorem}

\begin{prop}
    The weights of the linkage term in \textbf{Theorem \ref{thm:marker_marginal}} sum up to a value smaller than $1$.
    
    \begin{align*}
        \frac{\E_{\bS}[\Var(M_{i1} \mid \bS_i)]}{\Var[\tM_{i1}]} \leq 1
    \end{align*}
    where the equality holds if and only if $\E[M_{i1} \mid \bM_{i(-1)}] = \lp[M_{i1} \mid \bM_{i(-1)}]$.
    \label{prop:general_wahlund}
\end{prop}

\begin{theorem}
    Let $\bM_{i(-1)}^{(q)}$ be a sequence of markers other than $1$ with length $q$.
    Assume that $\bM_{i(-1)}^{(q)}$ defines a monotonically increasing filteration $\sigma  \langle \bM_{i(-1)}^{(q)} \rangle$ respect to $q$.
    Under the casual structure of section (\ref{subsec:cstruct}),
    
    \begin{align*}
        \E_{\bS}[\E(M_{i1} \mid \bS_i) \mid \bM_{i(-1)}] \xrightarrow[q \rightarrow \infty]{}
        \E[M_{i1} \mid \bS_i =\bs] 
    \end{align*}
    in $L^1(\Pb)$.
    \label{thm:prefect_pred}
\end{theorem}

\textbf{Theorem \ref{thm:prefect_pred}} tells us that with sufficiently large number of markers as covariates, the prediction error becomes negligible.
This is very likely in modern GWAS since millions of variants are genotyped. 
We defer the detailed proof to the \textbf{Appendix} but provide a brief sketch here.

To prove the theorem, we resort ourselves to a more general problem where we consider the convergence of distributions, namely, $\Pb( \spdot \mid \bM_{i(-1)}) \rightarrow \Pb( \spdot \mid \bS_i=\bs)$ with $\bS_i =\bs$ fixed for an individual. 
If we view $\bs$ as the true parameter, $\Pb(\bS_i)$ as the prior distribution and $\Pb(\bS_i \mid \bM_{i(-1)})$ (where $\bM_{i(-1)}$ is the data) as the posterior distribution, the convergence is simply a \textit{posterior consistency} problem in Bayesian statistics \cite{Ghosal_2007, Ghosal_2017}.
Loosely speaking, posterior consistency means that the posterior distribution degenerates to a point mass as the number of data increases.
Hence, we evoke the Doob's theorem to prove our statement \cite{Ghosal_2017}.
The theorem tells us that we can exactly predict the population membership $\bS_i$ with large number of markers.

\begin{remark}
    The functional misspecification $\E[M_{i1} \mid \bM_{i(-1)}] - \lp[M_{i1} \mid \bM_{i(-1)}]$ is generally nonzero.
    One concrete example is the two discrete population model.
    A closed-form analysis can be delivered assuming that elements of $\bM_{i(-1)}$ are mutually unlinked. 
    By the Bayes theorem,
    
    \begin{align*}
        \Pb ( \bS_i \mid \bM_{i(-1)} )
        = \frac{ \Pb( \bM_{i(-1)} \mid \bS_i) \; \Pb(\bS_i)}{ \Pb (\bM_{i(-1)})}
    \end{align*}
    This gives
    
    \begin{align*}
        \frac{
            \Pb ( \bS_i=1 \mid \bM_{i(-1)} )
        }{
            1- \Pb ( \bS_i=1 \mid \bM_{i(-1)} )
        }
        &= \frac{
            \Pb ( \bS_i=1 \mid \bM_{i(-1)} )
        }{
            \Pb ( \bS_i=0 \mid \bM_{i(-1)} )
        } \\
        &= \frac{
            \Pb( \bM_{i(-1)} \mid \bS_i=1) 
        }{
            \Pb( \bM_{i(-1)} \mid \bS_i=0)
        } \cdot
        \frac{
            \Pb(\bS_i=1)
        }{
            \Pb(\bS_i=0)
        }
    \end{align*}
    On the other hand, $\Pb(\bM_{i(-1)} \mid \bS_i)$ is factorized due to the conditional independence $M_{ik} \indep M_{ik'} \mid \bS_i$.
    
    \begin{align*}
        \Pb(\bM_{i(-1)} \mid \bS_i = \mathbf{s}) &= \prod_{k \neq 1} \Pb(M_{ik} \mid \bS_i)  \\
        &= \prod_{k \neq 1} \binom{2}{M_{ik}} \left(g_k^{\mathbf{s}} \right)^{M_{ik}} \left( 1- g_k^{\mathbf{s}} \right)^{2-M_{ik}} 
    \end{align*} 
    Finally, applying $\log( \spdot \; )$ on both sides of the equation shows that $\E[\bS_i\mid \bM_{i(-1)}]$ is a logit-linear function respect to $\bM_{i(-1)}$.
    
    \begin{align*}
        \mathrm{logit}(\E[\bS_i \mid M_{i(-1)}]) &=
        \sum_{k \neq 1} 
        \left[ 
        M_{ik} \log \left( \frac{g_k^1}{g_k^0} \right) + (2-M_{ik}) \log \left( \frac{1-g_k^1}{1-g_k^0} \right) \right] + \log \left( \frac{
            \Pb(\bS_i=1)
        }{
            \Pb(\bS_i=0)
        }\right)
    \end{align*}
    \label{rmk:func_mis}
\end{remark}

In sum, even with handful of covariates, the non-vanishing functional misspecification attenuates the true signal according to \textbf{Proposition \ref{prop:general_wahlund}} and leaves an additional confounding term according to \textbf{Theorem \ref{thm:marker_marginal}}.

\subsection{Within-sibship GWAS}

In within-sibship GWAS, we observe the family membership of each individual.
Denote the family membership of individual $i$ as $\bF_i$.
The families are index by $f$.

Marginal testing of markers are performed using the following regression.

\begin{align}
    Y_i = M_{i1} \beta_{1,\bf} + \sum_f \mathbf{1}(\bF_i=f)\gamma_f + \beta_0 + \delta_i
    \label{eq:marker_marginal_family}
\end{align}
The OLS estimator of regression (\ref{eq:marker_marginal_family}) is equivalent to the famous sibling difference regression when there are only two siblings  per family \cite{Brumpton_2020}.

\begin{align}
    Y_i - Y_{i'} = (M_{i1} - M_{i'1}) \beta_{1,\mathrm{sd}} + (\delta_i - \delta_{i'}) \text{\quad where\quad} \bF_i = \bF_{i'}
    \label{eq:marker_marginal_sd}
\end{align}
The proof of the algebraic equivalence can be found in Wooldrdige (2021) \cite{Wooldridge_2021}.

The estimand of regression (\ref{eq:marker_marginal_family}) is
\begin{align}
    \begin{aligned}
        \beta_{1,\bf} &= \sum_j \frac{\E_{\bF}[\beta_{1j}(\bF_i) \Var(M_{i1} \mid \bF_i)]}{\E_{\bF}[\Var(M_{i1} \mid \bF_i)]} \\
        &= \frac{\E_{\bF}[\beta_{1}(\bF_i) \Var(M_{i1} \mid \bF_i)]}{\E_{\bF}[\Var(M_{i1} \mid \bF_i)]}
    \end{aligned}
    \label{eq:marker_marginal_fam}
\end{align}
where $\beta_1(\bF_i=f)$ is the estimand of regression (\ref{eq:marker_marginal_family}) restricted to family $\bF_i = f$.
By combining the facts (1) regressions (\ref{eq:marker_marginal_family}) and (\ref{eq:marker_marginal_sd}) have algebraically identical OLS estimators and (2) OLS consistently estimates $\beta_{1,\bf}$, we show that the estimand $\beta_{1,\bf}$ is equal to $\beta_{1,\bs}$, the population estimand when $\bS_i$ is known.

\begin{theorem}
    If we ignore recombination, 
   \begin{align*}
       \frac{\E_{\bF}[\beta_{1}(\bF_i) \Var(M_{i1} \mid \bF_i)]}{\E_{\bF}[\Var(M_{i1} \mid \bF_i)]}
       = \frac{\E_{\bS}[\beta_{1}(\bS_i) \Var(M_{i1} \mid \bS_i)]}{\E_{\bS}[\Var(M_{i1} \mid \bS_i)]}
   \end{align*} 
   \label{thm:fam_pop_equivalence}
\end{theorem}

\subsection{Non-genetic confounding}
We initially assumed that $\E[ \epsilon_i \mid \bC_i, \bM_i] =0$ in the CAM model (\ref{eq:causal_whole}).
This can be relaxed so that $\epsilon_i = \nu_i + U_i$ where $\E[\nu_i \mid \bC_i, \bM_i] =0$ and $\E[U_i]=0$.
$U_i$ may contain non-genetic random variables.

Equation (\ref{eq:marker_marginal_s}) still holds provided 
\begin{enumerate}
    \item 
    \tikz[baseline]{
        \node[anchor=base] (g) {$\bM_i, \bC_i$}; 
        \node[anchor=base] (s) [right = of g] {$\bS_i$}; 
        \node[anchor=base] (u) [right = of s] {$U_i$}; 
        \path[->] (s) edge (g); 
        \path[->] (s) edge (u);
        }: 
        $U_i$ is correlated with the genetic variables only through $\bS_i$.
    \item $\E[U_i \mid \bS_i]$ is linear respect to $\bS_i$.
\end{enumerate}
which are identical to the conditions in which genetic confounding is resolved with linear regression.

The estimand using regression (\ref{eq:marker_marginal}) is also similar to \textbf{Theorem \ref{thm:marker_marginal}} with two additional terms.
\begin{align}
    \begin{aligned}
        &+ 
             \underbrace{\frac{
                \E[U_{i} ( \E[M_{i1} \mid \bS_i] - \E_{\bS}[\E(M_{i1} \mid \bS_i) \mid \bM_{i(-1)}])]
                }{
                    \Var[\tM_{i1}] 
                } 
            }_{\text{prediction error}} \\
            &+  \underbrace{\frac{
                \E[U_{i} ( \E[M_{i1} \mid \bM_{i(-1)}] - \lp[M_{i1} \mid \bM_{i(-1)}])] 
                }{
                    \Var[\tM_{i1}] 
                }
            }_{\text{functional misspecification}}
    \end{aligned}
    \label{eq:env}
\end{align}
Prediction error vanishes and the functional misspecification term is generally non-zero as before.
The derivation is essentially the same as in \textbf{Theorem \ref{thm:marker_marginal}} in which $C_{ij}$ is simply replaced by $U_i$.
Note that the result does not depend on the two conditions in the previous paragraph.
The formula shows how non-genetic confounding is partially resolved with genetic markers in linear regression: genetic markers indirectly controls the non-genetic variables by predicting $\bS_i$. 

\section{Conclusions}

In this work, we aimed to solve the identification problem of GWAS of quantitative traits using linear regression in a structured population consisting of subpopulations that are conditionally in HWE.
We established the connection between the CAM and the MAM which provides a closed-form formula for what is being estimated using linear regression.
The formula shows that under the popular population design, population structure exhibits a two-fold effect in which it induces an additive confounding term together with an attenuation of the true effect of a causal variant (\textbf{Proposition \ref{prop:marker_marginal_nocov}}).
As expected, within-sibship design can overcome this problem due to direct access to family membership.

The work shows how population structure is corrected in GWAS of quantitative traits.
By including unlinked markers to the regressions as covariates, linear regression implicitly removes the variation associated to the population structure from the variant being tested.
The remaining variation after the removal is then regressed on the trait which gives a less biased estimate.
Importantly, this means that the bias is corrected by modelling the distribution of the variant and not the trait as many have believed \cite{Sul_2018, McCaw_2022}.
Nevertheless, the bias is never completely removed because the expectation of the variant being tested is never truly linear respect to the covariate markers (\textbf{Theorem \ref{thm:marker_marginal}}).
In this viewpoint, it becomes clear how genetic covariates can further correct environmental confounding even without directly observing them as shown in equation (\ref{eq:env}).

We expect our framework to be extended to incorporate other important evolutionary processes such as assortative mating and inbreeding. 
As the (conditional) independence between the two haplotypes of an individual plays an important role in our results and proofs, haplotype dependence induced by such evolutionary processes is likely to have an non-trivial impact on GWAS estimands \cite{Veller_2023}.
Another shortcoming of our work is that it only deals with the identification and tells little about the estimation process.
Therefore, although our work deals with the biases in the estimate, it remains silent about the power and the precision of the estimators.

Among popular methods, only \textit{linear mixed models} (LMMs) exactly conform to equation (\ref{eq:marker_marginal}).
Since PC correction applies \textit{principal component analysis} (PCA) to $\bM_{i(-1)}$ prior to regression, it is a biased estimator for $\beta_1$ \cite{Mai_2022}. 
This fact, together with previous works on PCA that show that it asymptotically recovers admixture proportions, shows that a previously believed relationship between PC correction and LMMs is incorrect \cite{McVean_2009, Zheng_2016}.
Previous work had argued that LMMs confer a stronger correction against population structure compared to PC correction \cite{Vilhj_lmsson_2012,Hoffman_2013}.

Two distinct arguments show that this claim is wrong.
First, because PCA recovers the admixture proportion asymptotically, including the PC covariates properly adjusts population structure with the estimand in equation (\ref{eq:marker_marginal_s}).
Meanwhile, LMM fails to do so because it conforms to the MAM that leads to \textbf{Theorem \ref{thm:marker_marginal}}.
Second, an important feature of LMMs, that was missed by previous analyses based on the number of eigen components considered, is that random effects are assumed to be independent from the variant being tested.
Such an assumption is equivalent to the claim that the variant of interest is not subject to population structure which makes the attempt to correct population structure obsolete.
This is a general shortcoming of random effects estimators that led to alternative methods in other fields \cite{Wooldridge_2010, Chamberlain_1982}.
On the other hand, PC correction includes covariates as fixed effects which does not assume such independence thereby making a correct adjustment.

\section*{Acknowledgements}
We thank the following colleagues who provided helpful feedbacks after reading an early version of the draft.
Our work would have been impossible without them.
Doc Edge (University of Southern California, US) gave important comments on a population genetic perspective.
Qingyuan Zhao (University of Cambridge, UK) suggested recent literature in statistics and probability related to our work.

\section*{Proofs}
\begin{proof}[Proof of \textbf{Proposition \ref{prop:marker_marginal_nocov}}]
It follows from \textbf{Theorem \ref{thm:marker_marginal}} by setting $\bM_{i(-1)}$ empty.
\end{proof}

\begin{proof}[Proof of \textbf{Theorem \ref{thm:marker_marginal}}]

By the Frisch-Waugh-Lovell (FWL) theorem \cite{Wooldridge_2010}, the estimand of regression (\ref{eq:marker_marginal}) is

\begin{align*}
    \beta_1 = 
    \frac{\E[Y_i \tM_{i1}]}{\E[\tM_{i1}^2]}
\end{align*}
and the estimand of the regression restricted to $\bS_i = \bs$ is

\begin{align*}
    \beta_1(\bS_i = \bs) 
    = \frac{\E[Y_i (M_{i1} -\E[M_{i1} \mid \bS_i =\bs])]}{\Var[M_{i1} \mid \bS_i]}
\end{align*}

Now we expand the numerator of $\beta_1$.
Substituting $Y_i$ with the CAM model (\ref{eq:causal_whole}) gives

\begin{align*}
    \begin{aligned}
        \E[Y_i \tM_{i1}] 
        &= \sum_j \E [ C_{ij} \tM_{i1} ] \alpha_j + \E[ \epsilon_i \tM_{i1}] \\
        &= \sum_j \E[C_{ij} \tM_{i1}]\alpha_j \\
        &= \sum_j \left( \E[C_{ij} \dM_{i1}] + \E[C_{ij} (\E[M_{i1} \mid \bS_i] - \lp[M_{i1} \mid \bM_{i(-1)})] \right) \cdot \alpha_j
    \end{aligned}
\end{align*}

The first term $\E[C_{ij} \dM_{i1}] \alpha_j$ can be expressed in terms of $\beta_{1j}(\bS_i = \bs)$ by substituting the expression of it in the first paragraph of the proof.

\begin{align*}
    \begin{aligned}
        \E[C_{ij} \dM_{i1}]\alpha_j &= \E_{\bS}[\E(C_{ij} \dM_{i1} \mid \bS_i)\alpha_j] \\
        &= \E_{\bS}[ \beta_{1j}(\bS_i) \Var(M_{i1} \mid \bS_i) ]
    \end{aligned}
\end{align*}

The second term is expanded by adding and subtracting $E[M_{i1} \mid \bM_{i(-1)}]$. 
\begin{align*}
    \begin{aligned}
        &\E[C_{ij} (\E[M_{i1} \mid \bS_i] - \lp[M_{i1} \mid \bM_{i(-1)})] \\
        &= \E[C_{ij}(\E[M_{i1} \mid \bS_i] - \E[M_{i1} \mid \bM_{i(-1)}])]
        + \E[C_{ij} (\E[M_{i1} \mid \bM_{i(-1)}] - \lp[M_{i1} \mid \bM_{i(-1)}])]
    \end{aligned}
\end{align*}

Finally, the following reformulation of $\E[M_{i1} \mid \bM_{i(-1)}]$ completes the proof.

\begin{align*}
    \begin{aligned}
        \E[M_{i1} \mid \bM_{i(-1)}] &= \E_{\bS}[\E(M_{i1} \mid \bM_{i(-1)},\bS_i) \mid \bM_{i(-1)}] \\
        &= \E_{\bS}[\E(M_{i1} \mid \bS_i) \mid \bM_{i(-1)}]
    \end{aligned}
\end{align*}
due to $M_{i1} \indep \bM_{i(-1)} \mid \bS_i$.

\end{proof}

\begin{proof}[Proof of \textbf{Proposition \ref{prop:general_wahlund}}]
Let $\dM_{i1} = M_{i1} - \E[M_{i1} \mid \bS_i]$.

\begin{align*}
    \begin{aligned}
        \frac{\E_{\bS}[\Var(M_{i1} \mid \bS_i)]}{\Var[\tM_{i1}]} 
        &= \frac{\E_{\bS}[\Var(M_{i1} \mid \bS_i)]}{\Var[\dM_{i1}]} 
        \cdot \frac{\Var[\dM_{i1}]}{\Var[\tM_{i1}]}
    \end{aligned}
\end{align*}
The first term is smaller than $1$ due to the law of total variance.

Now we show that the second term is smaller than $1$.

\begin{align*}
    \begin{aligned}
        \Var[ \dM_{i1}]&=
        \E\left[(M_{i1} - \E[M_{i1} \mid \bS_i])^2\right]  \\
        &=\E\left[(M_{i1} - \E[M_{i1} \mid \bS_i, \bM_{i(-1)}])^2\right] \\
        &\leq \E\left[(M_{i1} - \E[M_{i1} \mid \bM_{i(-1)}])^2 \right] \\
        &\leq \E\left[(M_{i1} - \lp[M_{i1} \mid \bM_{i(-1)}])^2\right] \\
        &= \Var[ \widetilde{M}_{i1} ]
    \end{aligned} 
\end{align*}
The second line follows from the conditional independence implied by the causal structure.
The third line follows from the property of conditional expectation.
The forth line follows from the fact that the conditional expectation minimizes the square norm.

The result can be informally inferred in a intuitive way using the causal graph.
Explaining the variance of $M_{i1}$ with $\bS_i$ is more effective than with $\bM_{i(-1)}$
because the path from $\bM_{i(-1)}$ from $M_{i1}$, $\bM_{i(-1)} \leftarrow \bS_i \rightarrow M_{i1}$, must go through the path $\bS_i \rightarrow M_{i1}$.

\end{proof}

\begin{proof}[Proof of \textbf{Theorem \ref{thm:prefect_pred}}]
See \textbf{Theorem 6.9} of Ghosal and van der Vaart \cite{Ghosal_2017} (Doob's theorem).
First, substitute the variables appearing in the theorem according to our notation.
Replace $X^{(n)}$ to $\bM_i^{(q)}$, $\sigma \langle X^{(1)}, X^{(2)}, \ldots \rangle$ to 
$\sigma \langle M_{i2}, M_{i3}, \ldots \rangle$, $\theta$ to $\bs$ and $\Pi$ to $\Pb(\bS_i)$.
Next, apply the theorem to $f(\bs)=\E[M_{i1} \mid \bS_i =\bs]$ which gives the desired result.
\end{proof}

\begin{proof}[Proof of \textbf{Theorem \ref{thm:fam_pop_equivalence}}]
The proof is based on Veller and Coop (2023) [cite].
For a fixed $\bS_i=\bs$, equation (7) of Veller and Coop shows that 
\begin{align*}
    \beta_{1,\mathbf{f}}(\bS_i = \bs) = \frac{2}{H_1(\bS_i=\bs)} \sum_j D_{j1}^{\bs} \alpha_j
\end{align*}
where $H_1(\bS_i=\bs) = 2g_1^{\bs}(1-g_1^{\bs})$ and $D_{j1}^{\bs} = h_{j1}^{\bs}-f_j^{\bs}g_1^{\bs}$.
Therefore, $\beta_{1,\mathbf{f}}(\bS_i = \bs) = \beta_1(\bS_i=\bs)$.
Note that $\lambda$ was replaced to $1$ and $l$ was replaced to $j$ to match our notation.
This implicitly assumes that families are nested within populations.
Finally, substituting the result to equation (\ref{eq:marker_marginal_family}) gives the desired result.

\end{proof}

\bibliographystyle{unsrt_short}

\end{document}